\documentclass[prl,
 reprint,
 superscriptaddress,
 amsmath,amssymb,
 aps
]{revtex4-2}

\usepackage[final]{pdfpages}
\usepackage{pgffor}
\makeatletter
\AtBeginDocument{\let\LS@rot\@undefined}
\makeatother
\usepackage{graphicx}
\usepackage{dcolumn}
\usepackage{bm}
\usepackage{physics}
\usepackage{xcolor}
\usepackage{soul}
\usepackage{natbib}
\usepackage{esint}
\usepackage{hyperref}
\usepackage[capitalise]{cleveref}
\usepackage[mathlines]{lineno}

\usepackage{xcolor,cancel}
\usepackage[normalem]{ulem}
\usepackage{orcidlink}

\hypersetup{
	colorlinks=true,      
	linkcolor=blue,       
	citecolor=blue,       
	filecolor=magenta,    
	urlcolor=blue         
}

\begin{document}

\title{Signatures of Quantum Phase Transitions in Driven Dissipative Spin Chains }

\author{Mostafa Ali\orcidlink{0009-0004-2277-0097}~}
\email{alimosta@msu.edu}
\affiliation{Department of Physics and Astronomy, Michigan State University, East Lansing, Michigan 48824 USA}

\author{Naushad A. Kamar\orcidlink{0000-0002-8600-615X}~}
\affiliation{Department of Physics and Astronomy, Michigan State University, East Lansing, Michigan 48824 USA}

\author{Alireza Seif\orcidlink{0000-0001-5419-5999}~}

\affiliation{Pritzker School of Molecular Engineering, University of Chicago, Chicago, IL 60637}

\author{Mohammad Maghrebi}
\affiliation{Department of Physics and Astronomy, Michigan State University, East Lansing, Michigan 48824 USA}
\affiliation{Pritzker School of Molecular Engineering, University of Chicago, Chicago, IL 60637}

\begin{abstract}
Open driven quantum systems have defined a powerful paradigm of nonequilibrium phases and phase transitions; however, quantum phase transitions are generically not expected in this setting due to the decohering effect of dissipation. In this Letter, we consider a quantum Ising model subject to bulk dissipation (at rate $\Gamma$) and show that, although the correlation length remains finite (hence no phase transition), it develops a pronounced peak close to the ground-state quantum critical point. 
While standard techniques fail in this regime, we develop a versatile analytical approach that becomes exact with vanishing dissipation ($\Gamma\to 0$ but finite $\Gamma t$).
On a technical level, our approach builds on previous work where the state of the system is described by a slowly evolving generalized Gibbs ensemble that accounts for the integrability of the Hamiltonian while treating dissipation perturbatively. Finally, we demonstrate a kind of \textit{universality} in that integrability-breaking perturbations of the Hamiltonian
lead to the same behavior. To this end, we first show that the steady state of a chaotic Ising Hamiltonian under local Markovian dissipation that preserves the Ising symmetry, and in the limit $\Gamma \to 0$, is identical to that of quench dynamics in the absence of dissipation. This intriguing connection then allows us to draw on recent findings about quantum phase transition signatures in quench dynamics.
\end{abstract}

\maketitle 
Driven-dissipative quantum systems provide an exciting platform to investigate nonequilibrium many-body physics.
Indeed, nonequilibrium steady states of driven-dissipative systems can host new phenomena and exotic states of matter that cannot be realized in their equilibrium counterparts \cite{Diehl_Micheli_Kantian_Kraus_Büchler_Zoller_2008,verstraete_quantum_2009}. 
These systems naturally emerge in a plethora of experimental platforms, including exciton-polariton fluids \cite{doi:10.1126/science.1074464, Kasprzak_Richard_Kundermann_Baas_Jeambrun_Keeling_Marchetti_Szymańska_André_Staehli_et_al.2006, PhysRevLett.96.230602, Byrnes_Kim_Yamamoto_2014, Rodriguez_Amo_Sagnes_LeGratiet_Galopin_Lemaître_Bloch_2016, PhysRevLett.118.247402}, trapped ions \cite{doi:10.1126/science.aad9958, Schindler_Müller_Nigg_Barreiro_Martinez_Hennrich_Monz_Diehl_Zoller_Blatt_2013}, Rydberg gases \cite{Peyronel_Firstenberg_Liang_Hofferberth_Gorshkov_Pohl_Lukin_Vuletić_2012, Firstenberg_Peyronel_Liang_Gorshkov_Lukin_Vuletić_2013, PhysRevLett.111.113901, PhysRevLett.113.023006}, and superconducting qubits \cite{Houck_Türeci_Koch_2012, PhysRevX.7.011016,PhysRevLett.132.010601}.  Furthermore, they can be efficiently realized in programmable quantum simulators \cite{Ma_Saxberg_Owens_Leung_Lu_Simon_Schuster_2019,doi:10.1126/science.adh9932,PRXQuantum.4.040329, Sierant2022dissipativefloquet,haack2023probing}. 

These advances notwithstanding, dissipation generically leads to mixing and a loss of quantum coherence, leading to an effective classical or thermal behavior \cite{PhysRevB.74.245316, PhysRevLett.97.236808, PhysRevX.10.011039,sieberer2023universality}. With the exception of a few (possibly fine-tuned) models \cite{Torre2010,Marino2016,Rota_2019,Znidaric_2010}, quantum phase transitions are generically not expected in driven-dissipative settings. 
Given the paradigmatic status of ground-state quantum phase transitions, it would be desirable to find their signatures in quantum simulators. While signatures of quantum phase transitions have been identified far from equilibrium \cite{Bhattacharyya2015,Roy2017,Surace_2020,Titum-2019, Titum_2020,Paul2024,Haldar2021,de2023non,Lesanovsky_2021,Joshi_2024,Hanai_2024}, dissipation, being unavoidable in quantum simulators, is expected to spoil those at long times.
On the theoretical side, a major challenge is that the analytical toolbox for open quantum systems is rather limited compared to the standard setting of condensed matter physics.
With dissipation taking the spotlight, in part due to the emergence of noisy quantum simulators, several exact results and exactly solvable models \cite{Kawabata2024,Zunkovic_2014, PhysRevA_89_052133,PhysRevLett.101.105701, PhysRevLett.106.217206,PhysRevLett.119.190402,Roberts_2023} have been discovered recently.
However, even for paradigmatic models such as a spin chain subject to loss in the bulk (with a nontrivial steady state), analytical solutions are simply scarce.

\begin{figure}[h]
    \centering
    \includegraphics[scale=0.4]{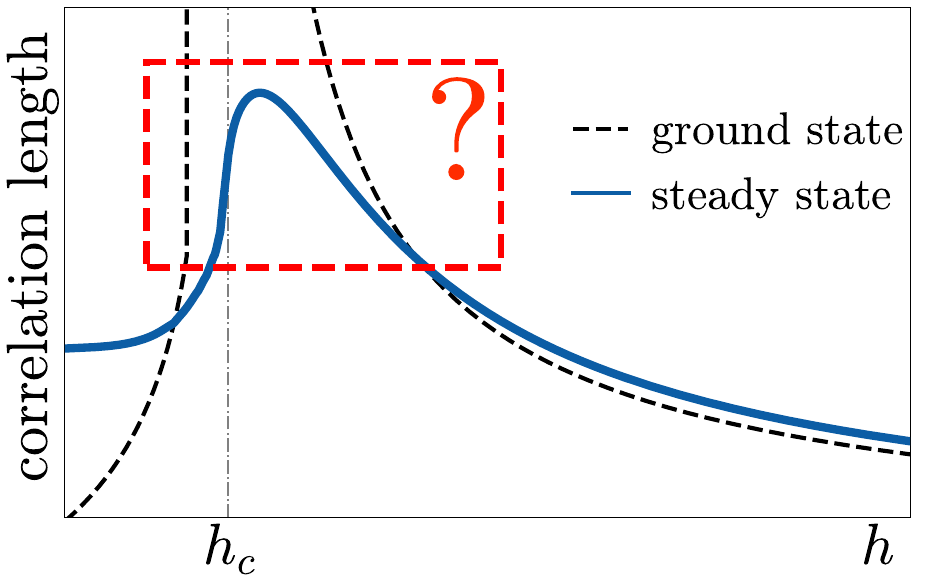}
    \caption{Schematic presentation of the correlation length in the steady state of a driven-dissipative Ising chain as a function of a tuning parameter $h$ (e.g., an external field), compared to that of the ground state. While being very distinct, the steady state still features a peak close to the ground-state quantum critical point $h_c$. An analytical approach is lacking in the regime highlighted by the question mark.}
    \label{fig:sketch correlation length}
\end{figure}

In this Letter, we investigate a quantum spin chain 
in an external field, $h$, and subject it to individual spin decay
[see \cref{eq:Liouvillian}]. 
In the absence of dissipation, this model undergoes a ground-state quantum phase transition, accompanied by a diverging correlation length, at a critical value, $h_c$. By contrast, the steady state of the driven-dissipative model is always disordered.
However, we show that, surprisingly, the correlation length peaks close to $h_c$; see \cref{fig:sketch correlation length}. 
With the exception of extreme limits $h\to 0, \infty$, a deeper, analytical understanding is simply unavailable. 
Even when the Hamiltonian dynamics can be mapped to free fermions, the addition of dissipation renders the model highly nonlinear. A naive approximation by simply dropping the nonlinear terms does not even correctly capture the extreme limits. 
Here, we develop an alternative analytical approach, inspired by the treatment of Bose gases in Ref.~\cite{10.21468/SciPostPhys.9.4.044}, in the limit of weak dissipation. In this limit, the system can be described by a generalized Gibbs ensemble (GGE) that accounts for the (free-fermion) integrability of the Hamiltonian, while treating the dissipation perturbatively \cite{Lange_Lenarcic_Rosch_2017, PhysRevB.97.165138, 10.21468/SciPostPhys.9.4.044,Lesanovsky_2023_1,Lesanovsky_2023_2,Dubail_2024,Dubail_2021,Barbier_2021,Perfetto_2024,lenarcic_2024,abanin_2024}. Remarkably, this approach closely matches the results obtained from numerical simulations at arbitrary $h$ and small but finite $\Gamma$. Finally, we consider integrability-breaking perturbations of the Hamiltonian (beside dissipation) and show that the peak moves even closer to the corresponding quantum critical point (QCP), hinting at a kind of \textit{universality}. In the limit $\Gamma \to 0$, we establish that the steady state is identical to that of a quantum quench, and present a generalization of the GGE ansatz to systems with chaotic Hamiltonians. We elucidate the generic sensitivity to the QCP by drawing on recent findings in the studies of quantum quenches \cite{Titum-2019,Haldar2021}.

\textit{Model---}
Here, we consider a driven-dissipative quantum Ising model where the Hamiltonian is given by
\begin{equation}
    \hat H = \sum_{i=1}^L -\sigma_i^x \sigma_{i+1}^x + h \sigma_i^z
    ,\label{eq:Ham}
\end{equation}
where $\sigma_i^{x,y,z}$ denote the Pauli operators and $h>0$ is the transverse field (with open boundary conditions, $\sigma_{L+1}^x=0$, for now).
Before considering dissipation, we remark that the ground state of this Hamiltonian makes a transition from a paramagnetic phase for $h>1$ to a ferromagnetic phase at $h_c=1$.
As shown in \cref{fig:sketch correlation length}, the phase transition is characterized by a diverging correlation length $\xi$ associated with spin-spin correlations,
$\langle\sigma_i^x \sigma_j^x\rangle \propto \exp \frac{-\abs{i-j}}{\xi}$. 
In this work, we assume that
individual spins decay at a rate $\Gamma$ as described by the Lindblad operator $\hat L_i=\sqrt{\Gamma} \sigma_i^-$. 
The full dynamics of the system's density matrix $\hat\rho$ is then governed by the quantum master equation
\begin{equation}
\begin{split}
    &\dv{\hat \rho}{t}=-i\comm{\hat H}{\hat \rho}+\sum_{i=1}^L {\cal D}_i(\hat \rho), \\
    &{\cal D}_i(\hat \rho) \equiv \hat L_i \hat \rho \hat L_i^\dagger - \frac{1}{2} \acomm{\hat L_i^\dagger \hat L_i}{\hat \rho}
    .\label{eq:Liouvillian}
\end{split}
\end{equation}
Such models have been proposed in several quantum simulation platforms, including superconducting circuits \cite{PhysRevLett.132.010601}, trapped ions \cite{Sierant2022dissipativefloquet,haack2023probing}, Rydberg atoms \cite{PhysRevA.95.042133}, and an array of coupled cavities \cite{PhysRevLett.109.130402,PhysRevLett.122.043602, PhysRevA.94.033801}. Unlike the ground state, the steady state of the driven-dissipative model does not host a phase transition as the quantum coherence is destroyed due to Markovian loss \cite{PhysRevB.93.014307, PhysRevA.95.042133,PhysRevA.88.063835,PhysRevLett.122.043602}. Consequently, the correlation length always remains finite; however, as we discuss shortly, a pronounced feature in the \textit{steady state} appears near the critical point of the \textit{ground-state} phase transition. 

\begin{figure}
        \centering
        \includegraphics[scale=0.43]{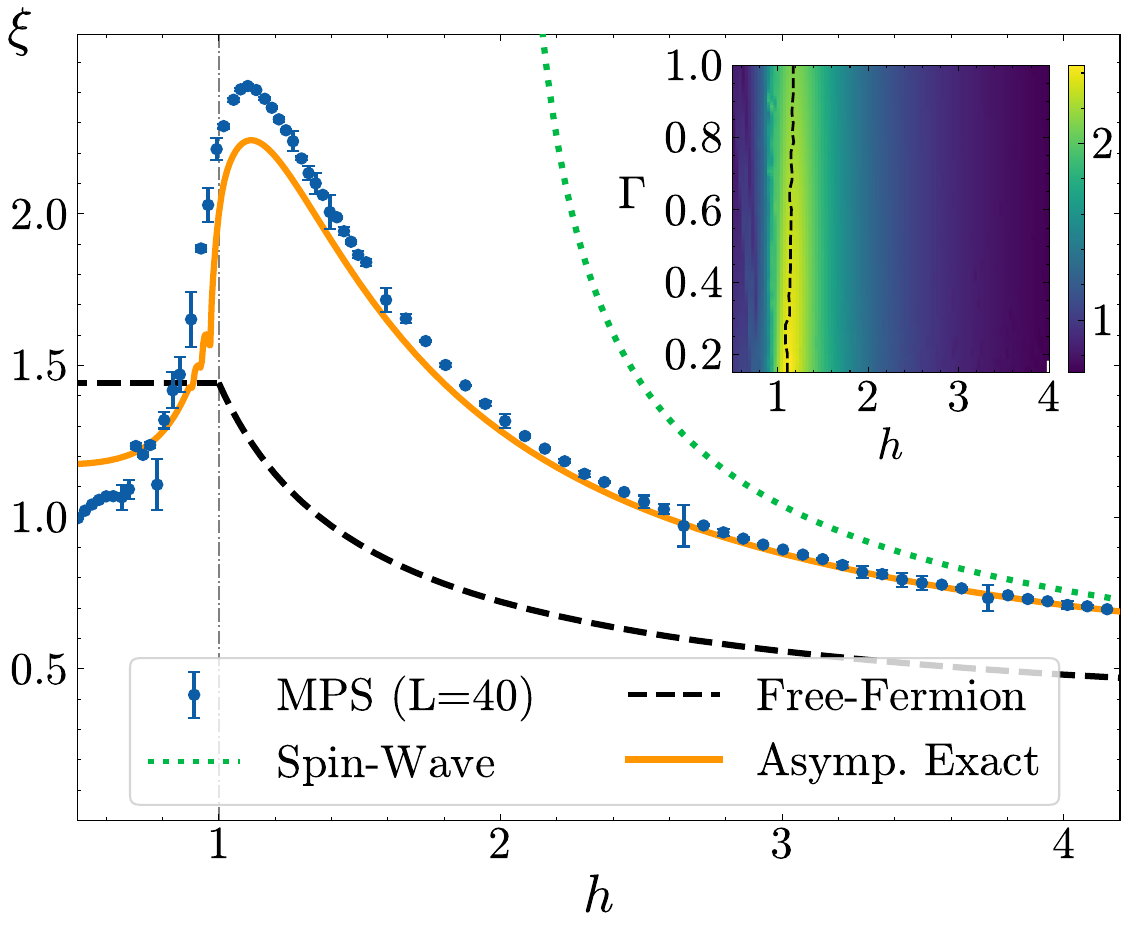}
        \caption{Correlation length $\xi$ calculated in the steady state of the driven-dissipative Ising model. MPS, spin-wave theory, and the free-fermion jump calculations are done at $\Gamma=0.15$. The solution that we developed, plotted in solid line, is asymptotically exact in the limit $\Gamma \to 0$. The vertical dashed line marks the QCP. In the inset, we present extensive MPS calculations for different $h$ and $\Gamma$ and for $L=40$ spins ($L=50$ does not result in a significant change). The dashed line indicates the position of the maximum correlation length for a given value of $\Gamma$. The correlation length decays as $\Gamma$ increases, but the peak remains close to the critical point. }
        \label{fig:born vs MPS}
    \end{figure}

We first present the results of an exact numerical simulation based on a variational matrix product state (MPS) approach 
\cite{PhysRevA.92.022116} combined with a split-basis local Hilbert space \cite{kamar2023splitting, PhysRevResearch.5.023026}; see Supplemental Material (SM) \cite{sm} for more details.
The correlation length for a chain of $40$ spins is presented in \cref{fig:born vs MPS} at weak dissipation. 
While the correlation length $\xi$ is always finite, 
we observe that it peaks at $h_{\rm peak}\approx1$ close to the ground-state critical point. Remarkably, this peak persists for a wide range of $\Gamma$; see the inset. 
Moreover, we find that $\xi$ in the steady state is larger than that in the ground state sufficiently away from the critical point \cite{sm}; see the schematic representation in \cref{fig:sketch correlation length}.
Short-range correlations were numerically observed to exhibit a peak too 
\cite{PhysRevA.88.063835, PhysRevA.92.022116}; however, a proper understanding is simply lacking. This is in part because no analytical treatment is available in the regime where $h\sim 1$. A simple spin-wave analysis or a mapping to free fermions (while exact in the absence of dissipation) simply fails in this regime. Prior to presenting our analytical approach, we briefly overview these simplistic methods. 

\textit{Spin-wave analysis---}
As a first attempt, we perform spin-wave analysis by mapping spins to bosons under the Holstein-Primakoff approximation $\sigma_i^- \to \hat b_i,\ \sigma_i^+ \to \hat b_i^\dag$, and $\sigma_i^z \to 2 \hat b_i^\dag \hat b_i -1$, where $\hat b_i\, (\hat b_i^\dag)$ is the bosonic annihilation (creation) operator. This approximation is expected to hold where the bosonic occupation is low, $\langle\hat b_i^\dag \hat b_i\rangle\ll1$. To solve for the steady state of the bosonic model, we utilize the Heisenberg-Langevin equation 
\cite{PhysRevA.88.063835}, taking the thermodynamic limit $L\to\infty$, and assuming periodic boundary conditions; see SM \cite{sm}. For $h \gg 1$, we obtain $\xi \sim 1 /\log{h}$ in good agreement with the numerical results; see \cref{fig:born vs MPS}. 
On the other hand, spin-wave analysis predicts that $\xi \to \infty$ as $h\to 2$ in the limit $\Gamma \to 0$ and breaks down when $h\le 2$, and thus it fails to capture the physics except for large $h$.

    \textit{Free fermions---}
    Another approximation is motivated by the fact that the Hamiltonian maps to free fermions under the Jordan-Wigner (JW) transformation, $\sigma^-_j=e^{i\pi \sum_{l=1}^{j-1} \hat c_l^\dagger \hat c_l}\hat c_j^\dagger$ and $\sigma^z_j=1-2\hat c_j^\dagger \hat c_j$, where $\hat c_j\, (\hat c_j^\dagger)$  is the fermionic annihilation (creation) operator; see \cref{eq:H_e/o}. However, applying the same transformation to the dissipative terms (specifically the jump term $\hat L_i \hat\rho \hat L_i^\dagger $) generates long string operators. A naive approximation would be simply to drop the string operator, and write the Lindblad jump operators as $\hat L_i=\sqrt{\Gamma}\sigma_i^- \to \sqrt{\Gamma}\hat c_i^\dagger$ in \cref{eq:Liouvillian}. 
    In the thermodynamic limit, the solution for this model in the steady state is also depicted in \cref{fig:born vs MPS}.
    Interestingly, the latter solution in the limit $\Gamma \to 0$
    coincides
    with the quench dynamics of the Ising model \cite{PhysRevA.69.053616}, in the \textit{absence} of dissipation, starting from an initial state with all spins pointing down \cite{sm}.
    The correlation length in this case is given by $\xi \sim 1/\log(2h)$ for $h\ge 1$, and $\xi =1/\log 2$ when $h<1$. 
    This approximation features a ``kink'' (that softens for finite $\Gamma$) in the correlation length at the quantum critical point; however, it still fails to capture the peak or even produce the correlation length for large $h$; see \cref{fig:born vs MPS}.

    \textit{Asymptotically exact solution---}
    The naive approximation that we discussed disregards the nonlocality of the spin operators by dropping the JW strings. In a full treatment, the \textit{parity} $e^{ i\pi\sum_{j=1}^L \hat c_j^\dagger \hat c_j}$ is not conserved: dissipation can remove a fermion hence changing the parity. Therefore, the  two parity sectors are coupled.
    Here, we assume that dissipation is weak compared to the Hamiltonian and perform a perturbative analysis which becomes asymptotically exact as $\Gamma\to 0$ but for any finite $\Gamma t$. Our approach is inspired by the treatment in Ref.~\cite{10.21468/SciPostPhys.9.4.044}, but is rather distinct since we consider an Ising chain as opposed to a Bose gas in the continuum (or hard-core bosons \cite{Dubail_2024}). We consider periodic boundary conditions, $\sigma^{\alpha}_{L+1}=\sigma^{\alpha}_1$, and take the thermodynamic limit at the end. For convenience, we also apply a $\pi$ rotation around the $x$ axis, taking $\sigma_{y,z}\to -\sigma_{y,z}$. Under the JW transformation, the Hamiltonian becomes block diagonal $\hat H = \hat H_{\rm e} \oplus \hat H_{\rm o} $ in the even and odd (e/o) parity sectors where
    \begin{equation}\label{eq:H_e/o}
        \begin{split}
            \hat H_{\rm e/o} = \sum_{j=1}^{L} (\hat c_j-\hat c^\dagger_j)(\hat  c_{j+1}+ \hat c^\dagger_{j+1}) +2h \hat c^\dagger_j \hat c_j,
        \end{split}
    \end{equation}
    with antiperiodic and periodic boundary conditions $\hat c_{L+1} = \mp \hat c_1$ for the even and odd sectors, respectively. 
    For each parity sector, this Hamiltonian can be diagonalized as $\hat H_{\rm e/o}=\sum_k \hat h_k\equiv \sum_k \epsilon_k \hat \alpha_k^\dagger \hat \alpha_k$ in terms of Bogoliubov fermions in momentum basis $\hat \alpha_k = \cos(\theta_k/2) \hat c_k -i\sin (\theta_k/2) \hat c_{-k}^\dagger$ where $\tan \theta_k= \frac{\sin k}{h-\cos k}$ and $\epsilon_k=2 \sqrt{1+h^2-2h\cos k}$. Note that, depending on the boundary conditions, momentum $k$ is quantized as $k \in \frac{2\pi}{L} \mathbb (\mathbb Z_L+\frac{1}{2}) \equiv \mathbb Z_{\rm ap}$, and  $k \in \frac{2\pi}{L} \mathbb Z_L  \equiv \mathbb Z_{\rm p}$ for even and odd parity sectors, respectively ($\mathbb Z_L = \mathbb Z \, {\rm mod}\, L$).
    In the absence of dissipation and given the integrability of the Hamiltonian, the system quickly relaxes to a state described by a generalized Gibbs ensemble (GGE) with an extensive number of conserved quantities \cite{Vidmar_2016}.
    
    Naively, one would take the system's density matrix as $\hat \rho_{\rm GGE}=\prod_k\hat\rho_k\sim e^{\sum_k\beta_k \hat h_k}$, which takes the form of a Gaussian operator with $\beta_k$ being the generalized inverse temperatures; however, 
    properly incorporating the parity, the density matrix becomes a block diagonal matrix consisting of two Gaussian operators \cite{10.21468/SciPostPhys.9.4.044}:
    \( 
        \hat \rho_{\text{{GGE}}} = 
        \prod_{k\in \mathbb{Z}_{\text{ap}}} \hat \rho_k \oplus
        \prod_{k\in \mathbb{Z}_{\text{p}}} \hat \rho_k
    \).
    Equivalently, the state can be fully characterized by the expectation value of the conserved charges $\hat Q_k \equiv \mathbb P_k \hat \alpha_k^\dagger \hat \alpha_k/2\pi$; here, $\mathbb P_k$ denotes a projector onto the even parity sector if $k\in \mathbb Z_{\rm ap}$ or the odd parity sector if $k\in \mathbb Z_{\rm p}$, respectively. With dissipation present but small, one can still describe the system in terms of a GGE 
    \cite{Lange_Lenarcic_Rosch_2017, PhysRevB.97.165138, 10.21468/SciPostPhys.9.4.044}; however, conservation becomes approximate and the charges $\langle \hat Q_k\rangle $ (or equivalently $\beta_k$) slowly evolve.
    Finally, using the GGE ansatz together with translation invariance, the dynamics for $Q_k \equiv \langle \hat Q_k\rangle$ is given by
    \begin{equation}\label{eq:Born_approx}
        \dv{  Q_k}{t} = \sum_{i=1}^L\big\langle {\cal D}_i^{\dagger}( \hat Q_k)\big\rangle_{\rm GGE} = L \big\langle {\cal D}_1^{\cal \dagger}(\hat Q_k) \big\rangle_{\rm GGE},
    \end{equation}
    where the adjoint superoperator ${\cal D}^\dagger$ governs the dynamics in the Heisenberg picture and the last equality follows from translation invariance. 
    In the above equation, the expectation value is taken with respect to the GGE state, hence with no contribution from the Hamiltonian. 
    With the dissipator now acting on the first site only, we can simply make the substitution $\hat L_1 = \sqrt{\Gamma}\sigma_1^- \to \sqrt{\Gamma} \hat c_1^\dagger$ without the JW string, significantly simplifying the subsequent calculations. We can then use Wick's theorem to compute the rhs of this equation. The anticommutator term $\{\hat c_1^\dagger  \hat c_1,\hat Q_k\}$ is easy to treat since it does not mix the parities; however, the jump term $\hat c_1^\dag \hat Q_k  \hat c_1$ changes the parity and thus the boundary conditions. Computing the latter term then requires a change of the basis 
    \cite{Dubail_2024}.
    To this end, we note that either periodic ($\ell \in \mathbb Z_{\rm p}$) or antiperiodic functions ($k\in \mathbb Z_{\rm ap}$) form a complete basis and are related to each other by \cite{Iorgov2011}
    \begin{equation}
        \hat c_k = -\frac{2}{L}\sum_{\ell \in \mathbb Z_{\rm p}} \frac{1}{1-e^{i(k - \ell)}} \hat c_{\ell}
        .
    \end{equation}
    In principle, 
    the application of  Wick's theorem together with the basis transformation allows us to compute the rhs of \cref{eq:Born_approx}. The resulting equations take a simpler form in terms of the original fermions $\hat c_k, \hat c^\dagger_k$ rather than Bogoliubov fermions $\hat \alpha_k, \hat \alpha_k^\dagger$.
    Indeed, the system can be fully characterized by the expectation value of the operators 
    $ \hat A_k=  \mathbb P_k \hat c_k^\dagger \hat c_k/2\pi$  and $i{ \hat B}_k =\mathbb P_k \hat c_k \hat c_{-k}/2\pi$. Ultimately, we aim to compute the corresponding expectation values $A_k, B_k$ (denoted by $c$ numbers with no hat). The functions $A_k,B_k$ are not independent as there is only a single conserved charge per $k$: the GGE constrains them as
    \begin{equation}     \label{eq:constraint}
        C_k \equiv ( A_k  - \frac{1}{4 \pi})\sin{k} -(h - \cos{k})  B_k  = 0.
    \end{equation}
    For operators $A$ ($B$), not conserved under the Hamiltonian dynamics, 
    the full dynamics is then subject to the above constraint:
    ${\rm d}A_k /{\rm d}t = -F_A(k)+\Lambda_k\partial C_k/\partial A_k$ (and similarly for $B$) where $\Lambda_k$ defines a Lagrange multiplier, and the terms $F_{A,B}$ are computed
    from \cref{eq:Born_approx} upon the substitution $\hat Q \to \hat A, \hat B$, respectively. Applying Wick's theorem then yields a nonlinear equation for $A_k, B_k$ which is also nonlocal in momentum (due to the basis transformation). In the thermodynamic limit $L\to\infty$, this equation can be written in a compact form by introducing the Hilbert transform $\cal H$ on the unit circle defined as $({\cal H} f)_k =\frac{1}{2\pi}{\rm P}\int_0^{2\pi} d\lambda f_\lambda\cot(\frac{k-\lambda}{2})$ with $\rm P$ being the principal part. The resulting equations take the form~\cite{sm}
        \begin{align}
        \begin{split}
            F_A (k) =& A + \frac{n^2}{2\pi}- 2\pi \left[A^2- ({\cal H}A)^2 - B^2+ ({\cal H}B)^2 \right]\\
             &+ 2n  ({\cal H}A)' ,\\
            F_B (k)=& B- 4\pi[AB - ({\cal H}A) ({\cal H}B)] +2n  ({\cal H}B)', 
        \end{split}
        \label{eq:solution_in_k}
    \end{align}
    where $n=\int \frac{dk}{2\pi}\langle \hat c_k^\dagger \hat c_k \rangle$ is the fermionic density and the prime indicates partial derivative with respect to $k$ which itself is implicit on the rhs.
    The resulting integro-differential equations take a simpler form once analytically continued into the complex plane (see SM \cite{sm}). Together with \cref{eq:constraint}, these equations allow us to solve for the fermionic correlations $A_k,B_k$. We can then determine spin correlations by computing the determinant of a matrix consisting of real-space fermionic correlators, and finally extract the correlation length $\xi$; see SM \cite{sm} for details. 
    
    In \cref{fig:born vs MPS}, we present the correlation length obtained from our analytical approach. Our results 
    closely match the MPS simulations and clearly reproduce a peak close to, but slightly above, the quantum critical point at
    \(
        h_{\rm peak} \approx 1.11
    \)
    in the limit $\Gamma\to 0$, thus confirming that $h_{\rm peak}\gtrapprox 1$ 
    is not an artifact of numerics at finite $L$ or $\Gamma$. Evidently, careful treatment of the boundary term in the Hamiltonian couples the two fermionic parity sectors, manifesting in the nonlinear terms in \cref{eq:solution_in_k}. This parity mixing is crucial for the emergent peak which is otherwise absent under the free-fermion approximation. Furthermore, our solution accurately predicts the spin correlations for $h<1$; in this regime, correlations do not simply decay exponentially but also exhibit oscillating spatial features.

    \begin{figure}[t]
        \centering
        \includegraphics[scale=0.44]{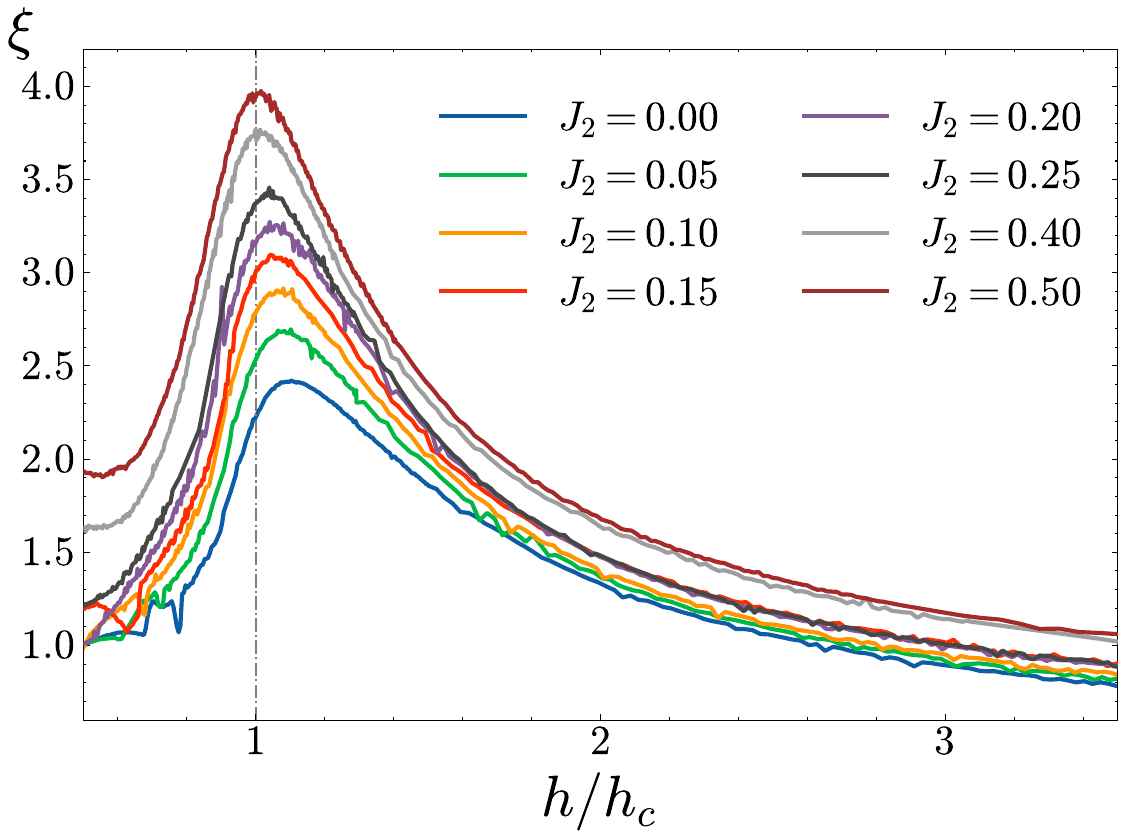}
        \caption{Correlation length $\xi$ computed in the steady state of the driven-dissipative NNN Ising model ($\Gamma=0.15$) obtained from MPS numerical results. The corresponding QCPs for $0\leq J_2 \leq 0.5$ are at $1\leq h_c \lessapprox 1.78$}
        \label{fig:NNN MPS}
    \end{figure}

    \textit{Beyond integrable Hamiltonians---}%
    An integrable (free-fermion) Hamiltonian allows for an analytical treatment; however, even when considering integrability breaking perturbations, the steady state remains remarkably sensitive to the quantum critical point. Here, we consider a model with ferromagnetic next-to-nearest-neighbor interactions,
    \begin{equation}
        \hat H_2 = \sum_{i=1}^L -\sigma_i^x \sigma_{i+1}^x - J_2 \sigma_i^x \sigma_{i+2}^x + h \sigma_i^z
        .\label{eq:Ham_NNN}
    \end{equation}
    
    We first note that this model undergoes a similar ground-state phase transition at $h_c(J_2)>1$ when $J_2>0$. In \cref{fig:NNN MPS}, we depict the correlation length in the steady state obtained from the MPS simulation of a chain of 40 spins. We observe that, the correlation length still peaks near the corresponding QCPs. In fact, the peak moves closer and closer toward $h/h_c=1$ as $J_2$ increases. While a phase transition is absent in our models, our observations suggest a form of \textit{universality}, where pronounced features appear even closer to criticality despite increasing integrability breaking perturbations. Thus, the sensitivity to the QCP is not tied to integrability, and should be understood in a more general setting.
    
    As a first step, we highlight several facts that establish an intriguing connection between quench dynamics of isolated, but chaotic, systems and driven-dissipative dynamics:
    ({1})~It is well established that the evolution under a chaotic Hamiltonian drives an isolated system (in the absence of dissipation) toward thermal equilibrium \cite{rigol2008thermalization}.
    ({2})~Recent work \cite{mori_2020,altman2020a,podolsky2018a,rosch2018c,rosch2018d} has shown that, under a chaotic Hamiltonian and bulk dissipation, driven-dissipative systems are also described by a thermal Gibbs ensemble (GE) in the limit $\Gamma \to 0$ (after taking $t\to\infty$); this extends our treatment of integrable Hamiltonians via the GGE ansatz.
    While neither ({1},{2}) involves an intrinsic temperature (e.g., due to an external reservoir), an effective temperature is set by the energy density. 
    ({3}) Finally, we show that, for a general Ising Hamiltonian subject to our choice of Markovian dissipation, the steady state's energy density is simply  
    $-h$ independent of other energy scales \cite{sm}, and identical to that of a fully polarized state $\ket{\psi_0} = \ket{\downarrow \downarrow ... \downarrow}$. Combining ({1}-{3}), we arrive at the nontrivial conclusion that the steady state of driven-dissipative dynamics under a chaotic Ising Hamiltonian in the limit $\Gamma\to 0$ is identical to that of quench dynamics under the same Hamiltonian and starting from $\ket{\psi_0}$. More generally, a quench protocol can be shown to arrive at the driven-dissipative steady state when the dissipation is local and respects the $\mathbb{Z}_2$ symmetry of the Ising Hamiltonian (see SM for details \cite{sm}).   

    Having established the above connection, we can gain some insight by examining quench dynamics near the QCP. For integrable systems, extensive work \cite{Bhattacharyya2015,Roy2017,Titum-2019,Titum_2020,Paul2024} has shown that quench dynamics is sensitive to the QCP, a behavior that can be attributed to the (single-particle) dispersion relation. On the other hand, it is not at all obvious if chaotic systems should be sensitive to the QCP. Interestingly however, signatures of quantum phase transitions have been recently identified in quench dynamics of chaotic Ising-type models \cite{Titum-2019,Haldar2021}. More precisely, when quenched from a fixed initial state (such as $|\psi_0\rangle$) to the vicinity of the QCP, spin correlators develop prominent peaks  \cite{Haldar2021}. Using the above analogy, Ising models subject to local Markovian dissipation should exhibit distinctive features near the QCP, which is exactly what we observe. 
    While a complete understanding of the observations in Refs.~\cite{Titum-2019,Haldar2021} is still open ended (see, e.g., \cite{Essler_2023}), they uncover an unexpected link between quantum phase transitions and quench dynamics, with immediate consequences for weakly dissipative open quantum chains under local Markovian dissipation.

    Before concluding, we point out that, despite the integrability breaking perturbation in the Hamiltonian (\ref{eq:Ham_NNN}), the corresponding steady state still deviates from a Gibbs ensemble. Beside finite-size effects, this deviation may be explained by the presence of long-lived quasi-conserved modes for small values of $J_2$ \cite{rosch2018c,lenarcic2021}.

    \textit{Conclusion---}%
    In this Letter, we have studied the intricate interplay of quantum phase transitions and dissipation in driven-dissipative spin chains. Specifically, we have shown that the \textit{steady state} of driven-dissipative Ising models exhibit a pronounced feature near the \textit{ground state} quantum critical point. For the nearest-neighbor Ising Hamiltonian, we have exploited (free-fermion) integrability as the basis for an asymptotically exact solution. Furthermore, we have observed the same features when integrability-breaking perturbations are included. To understand the case of nonintegrable Hamiltonians, we have established a nontrivial connection between driven-dissipative dynamics and chaotic dynamics in isolated quantum systems. On a conceptual level, our results are significant in that they show that quantum features could survive despite dissipation.
    On a practical level, our Letter opens the door to identifying quantum phase transitions in the context of noisy quantum simulators. Finally, on a technical level, our approach is immediately applicable to a large class of spin chains subject to decay in the bulk. The latter distinguishes our Letter from exact approaches focused on Hermitian jump operators (e.g., dephasing)  \cite{prosen2016a}, boundary dissipation \cite{Prosen_2015,Landi2022}, or collective models \cite{Bartolo2016,Roberts_2023}.

    \textit{Acknowledgments---}%
This Letter is supported by the Air Force Office of Scientific Research (AFOSR) under Award No. FA9550-20-1-0073. We also acknowledge support from the National Science Foundation under NSF CAREER Award No. DMR-2142866 as well as NSF Grant No. PHY-2112893. AS was partially supported by a Chicago Prize Postdoctoral Fellowship in Theoretical Quantum Science. The contributions of AS were completed while he was affiliated with the University of Chicago.

\bibliographystyle{apsrev4-2}

\newpage


\section*{Supplementary material}

# Supplemental Material for
# "Signatures of Quantum Phase Transitions in Driven Dissipative Spin Chains"

Mostafa Ali,[1] Naushad A. Kamar,[1] Alireza Seif,[2] and Mohammad Maghrebi[1]

[1]*Department of Physics and Astronomy, Michigan State University, East Lansing, Michigan 48824 USA*
[2]*Pritzker School of Molecular Engineering, University of Chicago, Chicago, IL 60637*

In this Supplemental Material, we provide additional details on the results stated in the main text. In Secs. S.I and S.II, we discuss the numerical solution provided in the main text and show the results obtained in a wide section of the parameter space. In Sec. S.III, we provide the details of the spin-wave theory calculations done under the Holstein-Primakoff approximation. In Sec. S.IV, we layout the asymptotically-exact treatment of the model under the assumption that the state of the driven dissipative Ising model assumes a GGE in the $\Gamma \to 0$ limit. We show how spin correlations can be recovered from the resulting fermionic correlations in Sec. S.V. In Sec. S.VI, we highlight the convergence of the free-fermion approximation to the results of quench dynamics in the limit $\Gamma \to 0$. In Sec. S.VII, we discuss the Binder cumulant calculations used to obtain the quantum critical point for non-integrable Hamiltonians.

## S.I. Numerical Solution

The numerical solution provided in the main text is based on a Matrix Product States (MPS) representation of the vectorized density matrix obtained under the mapping

$$\rho = \sum_{i,j} \rho_{ij} |i\rangle \langle j| \to |\rho\rangle\rangle = \sum_{i,j} \rho_{ij} |i\rangle \otimes |j\rangle . \tag{S.1}$$

The MPS ansatz takes the form

$$|\rho\rangle\rangle = \sum_{s} W[1]^{s'_1 s''_1} W[2]^{s'_2 s''_2} ... W[L]^{s'_L s''_L} |s'_1 s''_1, s'_2 s''_2 ..., s'_L s''_L\rangle , \tag{S.2}$$

where $W^{s'_i s''_i}$ is a rank-2 tensor of maximum dimensions $\chi^2 d^2$, where $d = 2$ is the local Hilbert space dimension, and $\chi$ is the MPS bond dimension. Furthermore, we split the local Hilbert space of $|\rho\rangle\rangle$, as described in [S1, S2], reducing the dimensionality of the tensors from $d^2$ to $d$. The final ansatz takes the form

$$|\rho\rangle\rangle = \sum_{s} A[1]^{s'_1} \tilde{A}[1]^{s''_1} A[2]^{s'_2} \tilde{A}[2]^{s''_2} ... A[L]^{s'_N} \tilde{A}[L]^{s''_L} |s'_1, s''_1, s'_2, s''_2 ..., s'_L, s''_L\rangle , \tag{S.3}$$

where the tensors $W[i]$ are now recast into the tensors $A[i]$, and $\tilde{A}[i]$. We then iteratively optimize each tensor $A[i](\tilde{A}[i])$ to the tensor corresponding to the zero eigenvalue of the effective Liouvillian operator acting on the site resulting in an approximate form of the steady state density matrix [S3].

## S.II. Correlation length in $(h, \Gamma)$ plane

Figure S.1 shows the correlation length obtained in the steady state of the driven-dissipative nearest-neighbor Ising model for a chain of length $L = 40$, for a range of $h$, and $\Gamma$ values. The peak of the correlation length is found to be close to the quantum critical point even for intermediate values of $\Gamma$.

## S.III. Spin-wave Theory

Here we provide the details for the spin-wave analysis reported in the main text. We find the steady state solution by applying the Heisenberg-Langevin equation [S4, S5] to the relevant operators. In the derivation, we consider a more general transverse-field Ising Hamiltonian, where $J_{m,n}$ is the Ising interaction only depending on the distance $|n - m|$,

$$\hat{H} = \sum_{n>m} J_{m,n} \sigma_m^x \sigma_n^x + h \sum_m \sigma_m^z . \tag{S.4}$$

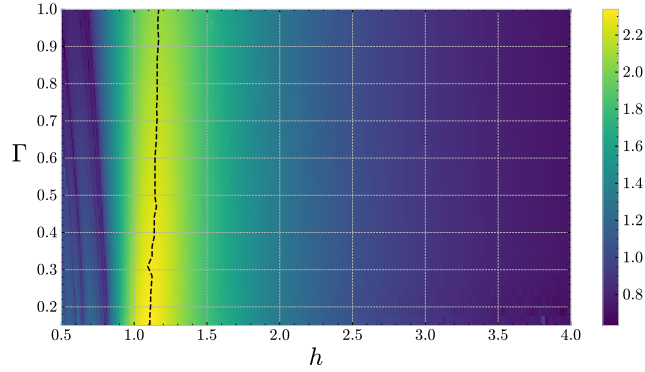

FIG. S.1. Correlation Length in the steady state of the driven dissipative nearest-neighbor Ising model against $h$ and $\Gamma$. Numerical results obtained from MPS calculations for $L = 40$. The dashed line indicates the position of the maximum correlation length for a given value of $\Gamma$. The peak slightly deviates away from the QCP as $\Gamma$ increases.

We apply a Holstein-Primakoff approximation under which we take

$$
\begin{aligned}
\sigma_m^- &\to \hat{b}_m, \\
\sigma_m^+ &\to \hat{b}_m^\dagger, \\
\sigma_m^x &\to \hat{b}_m + \hat{b}_m^\dagger, \\
\sigma_m^z &\to 2\hat{b}_m^\dagger \hat{b}_m - 1,
\end{aligned}
\tag{S.5}
$$

where $\left[\hat{b}_m, \hat{b}_n\right] = 0$, and $\left[\hat{b}_m, \hat{b}_n^\dagger\right] = \delta_{m,n}$. With that, we arrive at the following quadratic bosonic Hamiltonian

$$
\hat{H}_b = \sum_{n > m} J_{m,n} (\hat{b}_m + \hat{b}_m^\dagger)(\hat{b}_n + \hat{b}_n^\dagger) + h \sum_m (2\hat{b}_m^\dagger \hat{b}_m - 1).
\tag{S.6}
$$

Assuming periodic boundary conditions, we utilize the lattice translation-invariance symmetry to define the Fourier transformation $\hat{b}_m = \frac{1}{\sqrt{L}} \sum_k e^{-ikm} \hat{b}_k$, where the momentum modes $k$ take the values $k = \frac{2\pi l}{L}$ with $1 \le l \le L$. In Fourier space, momentum $k$ only couples to $-k$, and the Hamiltonian takes the simple form

$$
\hat{H}_b = \sum_k \hat{H}_k + \text{const},
\tag{S.7}
$$

where

$$
\hat{H}_k = \begin{pmatrix} \hat{b}_k^\dagger & \hat{b}_{-k} \end{pmatrix} \begin{pmatrix} J_k + h & J_k \\ J_k & J_k + h \end{pmatrix} \begin{pmatrix} \hat{b}_k \\ \hat{b}_{-k}^\dagger \end{pmatrix},
\tag{S.8}
$$

with $J_k \equiv \sum_{n-m>0} J_{m,n} \cos(k(n-m))$. Similarly, the dissipative part of the dynamics takes the form

$$
\mathcal{D}_b(\rho) = \Gamma \sum_k \hat{b}_k \rho \hat{b}_k^\dagger - \frac{1}{2} \hat{b}_k^\dagger \hat{b}_k \rho - \frac{1}{2} \rho \hat{b}_k^\dagger \hat{b}_k.
\tag{S.9}
$$

Under these conditions, the bosonic operators evolve in time under the Heisenberg-Langevin equation given by

$$
\begin{aligned}
\frac{d\hat{b}_k(t)}{dt} &= 2i\left[\hat{H}_k, \hat{b}_k\right] - \frac{\Gamma}{2}\hat{b}_k + \sqrt{\Gamma}\,\hat{b}_k^{in}(t), \\
\frac{d\hat{b}_{-k}^\dagger(t)}{dt} &= 2i\left[\hat{H}_k, \hat{b}_{-k}^\dagger(t)\right] - \frac{\Gamma}{2}\hat{b}_{-k}^\dagger(t) + \sqrt{\Gamma}\,\hat{b}_{-k}^{\dagger in}(t),
\end{aligned}
\tag{S.10}
$$

where we have used the fact that $\hat{H}_k = \hat{H}_{-k}$. $\hat{b}_k^{in}(t)$ are input noise operators satisfying the white-noise condition $\left\langle \hat{b}_k^{in}(t)\hat{b}_{k'}^{\dagger in}(t') \right\rangle = \delta_{k,k'}\delta(t - t')$. Other input noise correlators are identically zero. The above equations can be combined in the matrix form as

$$
\dot{f}(t) = \mathcal{M}f(t) + m(t),
\tag{S.11}
$$

where $f(t) = \begin{pmatrix} \hat{b}_k \\ \hat{b}_{-k}^\dagger \end{pmatrix}$, $m(t) = \begin{pmatrix} \sqrt{\Gamma}\hat{b}_k^{in} \\ \sqrt{\Gamma}\hat{b}_{-k}^{\dagger in} \end{pmatrix}$ and

$$\mathcal{M} = \begin{pmatrix} -2i(J_k + h) - \frac{\Gamma}{2} & -2iJ_k \\ 2iJ_k & 2i(J_k + h) - \frac{\Gamma}{2} \end{pmatrix}, \tag{S.12}$$

with the formal solution given by $f(t) = e^{\mathcal{M}t}f(0) + \int_0^t e^{\mathcal{M}(t-t')}m(t')dt'$. In the steady state, the first term vanishes, and we are left with the second term only. The matrix $e^{\mathcal{M}\tau}$ has the structure

$$e^{\mathcal{M}\tau} = \begin{pmatrix} g_1(\tau) & g_2(\tau) \\ g_2^*(\tau) & g_1^*(\tau) \end{pmatrix}. \tag{S.13}$$

Therefore, the solution in the limit $t \to \infty$ takes the form

$$f(t) = \sqrt{\Gamma} \int_0^t dt' \begin{pmatrix} g_1(t-t')\hat{b}_k^{in}(t') + g_2(t-t')\hat{b}_{-k}^{\dagger in}(t') \\ g_1^*(t-t')\hat{b}_{-k}^{\dagger in}(t') + g_2^*(t-t')\hat{b}_k^{in}(t') \end{pmatrix}. \tag{S.14}$$

Using the following definitions,

$$\begin{aligned} \epsilon_k &= 2(J_k + h), \\ \eta_k &= 2J_k, \\ \xi_k &= \sqrt{\epsilon_k^2 - \eta_k^2}, \end{aligned} \tag{S.15}$$

we can explicitly write the Green functions $g$ as

$$\begin{aligned} g_1(\tau) &= e^{-\frac{\Gamma}{2}\tau}[\cos(\tau\xi_k) - i\epsilon_k \frac{\sin(\tau\xi_k)}{\xi_k}], \\ g_2(\tau) &= -i\eta_k e^{-\frac{\Gamma}{2}\tau} \frac{\sin(\tau\xi_k)}{\xi_k}. \end{aligned} \tag{S.16}$$

Bosonic correlation functions in the steady state can thus be obtained using the solution in Eq. (S.14).

*Spin correlations.*—Under the HP approximation, the spin correlations at time $t$ at distance $l$ are given by

$$\langle \sigma_j^x(t)\sigma_{j+l}^x(t) \rangle = \langle \hat{b}_j^\dagger(t)\hat{b}_{j+l}(t) \rangle + \langle \hat{b}_j(t)\hat{b}_{j+l}(t) \rangle + \text{c.c.} \tag{S.17}$$

In the steady state ($t \to \infty$), they can easily be found by applying an inverse Fourier transformation to the bosonic correlators. The final results are given by

$$\langle \sigma_j^x \sigma_{j+l}^x \rangle_{SS} = \frac{1}{2\pi} \int_{-\pi}^\pi dk \ \cos kl \ \frac{\eta_k (\eta_k - \epsilon_k)}{(\frac{\Gamma}{2})^2 + \xi_k^2} + \frac{1}{2\pi} \int_{-\pi}^\pi dk \ \sin kl \ \frac{\eta_k (\frac{\Gamma}{2})}{(\frac{\Gamma}{2})^2 + \xi_k^2}. \tag{S.18}$$

As we showed in the main text, this result can be further simplified by analytically continuing $k$ to the unit disk in the complex plane ($e^{ik} \to z$) which gives Eq. (3).

## S.IV. Born approximation & GGE

For convenience, we first apply the transformation $\sigma_i^{y,z} \to -\sigma_i^{y,z}$. This transformation flips the sign of $h$ in the Hamiltonian

$$\hat{H} = -\sum_{i=1}^L \sigma_i^x \sigma_{i+1}^x + h\sigma_i^z, \tag{S.19}$$

and takes $\hat{L}_i = \sqrt{\Gamma}\sigma_i^- \to \sqrt{\Gamma}\sigma_i^+$.

### 1. Jordan-Wigner transformation

By performing a Jordan-Wigner transformation

$$\hat{\sigma}_i^- = \prod_{l=1}^{i-1}(1 - 2\hat{c}_l^\dagger \hat{c}_l)\hat{c}_i^\dagger, \quad \sigma_i^+ = (\sigma_i^-)^\dagger, \quad \sigma_i^z = 1 - 2\hat{c}_i^\dagger \hat{c}_i, \tag{S.20}$$

the Hamiltonian takes the form [S6]

$$\hat{H} = -\sum_{i=1}^{L-1}(\hat{c}_i^\dagger - \hat{c}_i)(\hat{c}_{i+1} + \hat{c}_{i+1}^\dagger) - h\sum_{i=1}^{L}(\hat{c}_i\hat{c}_i^\dagger - \hat{c}_i^\dagger\hat{c}_i) - e^{i\pi\hat{N}}(\hat{c}_L - \hat{c}_L^\dagger)(\hat{c}_1 + \hat{c}_1^\dagger), \tag{S.21}$$

where $\hat{N} = \sum_{i=1}^{L}\hat{c}_i^\dagger\hat{c}_i$ is the total fermionic number operators. Because the Hamiltonian is number-conserving, the even- and odd-parity sectors of $\hat{H}$ can be treated independently. One can write the Hamiltonian in even/odd sectors as

$$\hat{H}_{\text{e/o}} = -\sum_{i=1}^{L}(\hat{c}_i^\dagger - \hat{c}_i)(\hat{c}_{i+1} + \hat{c}_{i+1}^\dagger) - h\sum_{i=1}^{L}\hat{c}_i\hat{c}_i^\dagger - \hat{c}_i^\dagger\hat{c}_i, \tag{S.22}$$

$$\text{with} \quad \hat{c}_{L+1} = \mp\hat{c}_1.$$

That is, the Hamiltonian is that of free fermions but with anti-periodic/periodic boundary conditions for even/odd sectors, respectively. Next, one can diagonalize the Hamiltonian in terms of Bogoliubov fermions defined as [S6, S7]

$$\hat{\alpha}_k = \cos\left(\theta_k/2\right)\hat{c}_k - i\sin\left(\theta_k/2\right)\hat{c}_{-k}^\dagger, \tag{S.23}$$

where

$$e^{i\theta_k} = \frac{h - e^{-ik}}{\sqrt{1 + h^2 - 2h\cos k}}, \tag{S.24}$$

or $\tan\theta_k = \frac{\sin k}{h - \cos k}$ as stated in the main text. With this set of definitions, we find[1]

$$\hat{H}_{\text{e/o}} = \sum_{k \in \mathbb{Z}_{\text{ap,p}}} \epsilon_k[\hat{\alpha}_k^\dagger\hat{\alpha}_k - 1], \tag{S.25}$$

where with the dispersion relation

$$\epsilon_k = 2\sqrt{1 + h^2 - 2h\cos k}. \tag{S.26}$$

### 2. Basis transformation

For a closed system and for even observables (i.e., those not changing the parity), we can limit ourselves to the even sector [S6]; however, introducing dissipation violates parity conservation, and thus requires a change of basis states. Since both $c_k$ for $k \in \mathbb{Z}_{\text{ap}}$ and $c_p$ for $p \in \mathbb{Z}_{\text{p}}$ span the one-particle Hilbert space on a ring, they can be related to each other. Indeed, the corresponding momentum eigenstates are related via

$$|k\rangle = \sum_{p \in \mathbb{Z}_{\text{p}}} \langle p|k\rangle|p\rangle, \tag{S.27}$$

where

$$\langle p|k\rangle = \frac{1}{L}\sum_{j=1}^{L} e^{i(k-p)j} = -\frac{2}{L}\frac{1}{1 - e^{-i(k-p)}}, \tag{S.28}$$

---

[1] More precisely, defining the zero mode in the odd parity sector requires a particle-hole transformation for $h < 1$ [S6].

Identifying $|k\rangle = \hat{c}_k^\dagger |0\rangle$, we find

$$\hat{c}_k = -\frac{2}{L} \sum_{p \in \mathbb{Z}_p} \frac{1}{1 - e^{i(k-p)}} \hat{c}_p, \tag{S.29}$$

Up to a relative sign, this expression is given in Ref. [S8]. Taking the thermodynamic limit $L \to \infty$, we recover the basis transformation $\hat{c}_k = -\frac{2i}{L} \sum_{p_m \in \mathbb{Z}_p} \frac{1}{k_n - p_m} \hat{c}_p$ reported in [S9]. Finally, from the completeness relation

$$\sum_{p \in \mathbb{Z}_p} \langle k'|p \rangle \langle p|k \rangle = \delta_{k,k'}, \tag{S.30}$$

we arrive at the condition

$$\sum_{p \in \mathbb{Z}_p} \frac{1}{1 - e^{-i(k-p)}} \frac{1}{1 - e^{i(k'-p)}} = \frac{L^2}{4} \delta_{k,k'}. \tag{S.31}$$

### 3. Wick's theorem

As stated in the main text, it is more convenient to work with the $c, c^\dagger$ fermions (rather than Bogoliubov fermions). To this end, we observe that $A_k = \langle \hat{c}_k^\dagger \hat{c}_k \rangle / 2\pi$ and $B_k \equiv \langle \hat{c}_k \hat{c}_{-k} \rangle / 2\pi i$ are constrained together as the only nontrivial correlator is given by $\langle \hat{\alpha}_k^\dagger \hat{\alpha}_k \rangle$. Inverting Eq. (S.23), we obtain the expectation values

$$2\pi A_k = \frac{1}{2} + \cos\theta_k p_k, \qquad 2\pi B_k = \sin\theta_k p_k, \tag{S.32}$$

where $\langle \alpha_k^\dagger \alpha_k \rangle = \frac{1}{2} + p_k$. We thus find

$$(A_k - 1/4\pi)\sin\theta_k - B_k\cos\theta_k = 0, \tag{S.33}$$

which reduces to Eq. (7) of the main text, and is repeated here for completeness:

$$(A_k - 1/4\pi)\sin k - B_k(h - \cos k) = 0. \tag{S.34}$$

We treat the constraint which is linear in $A_k, B_k$ via Lagrange multipliers. Combined with the Born approximation, this yields

$$\frac{1}{\Gamma}\frac{\mathrm{d}}{\mathrm{d}t} A_k = -F_A(k) + \Lambda_k \sin k, \tag{S.35a}$$

$$\frac{1}{\Gamma}\frac{\mathrm{d}}{\mathrm{d}t} B_k = -F_B(k) - \Lambda_k (h - \cos k), \tag{S.35b}$$

where we have defined

$$-\frac{1}{L} F_A \equiv \langle \hat{c}_1^\dagger \hat{A}_k \hat{c}_1 \rangle - \langle \hat{c}_1^\dagger \hat{c}_1 \hat{A}_k \rangle, \tag{S.36}$$

and similarly for $F_B$; here the expectation value is computed with respect to the GGE state (the label dropped for notational ease). Now, the last term in the above equation does not involve a change of parity ($\hat{c}_1^\dagger \hat{c}_1$ conserves the particle number), and can be easily computed by noting $\hat{c}_1 = \frac{1}{\sqrt{L}} \sum_\lambda e^{ik} \hat{c}_k$ and using Wick's theorem as

$$\frac{L}{2\pi} \langle \mathbb{P}_k \hat{c}_1^\dagger \hat{c}_1 \hat{c}_k^\dagger \hat{c}_k \rangle = nLA_k + A_k(1 - 2\pi A_k) + 2\pi B_k^2. \tag{S.37}$$

On the other hand, the first term on the rhs of Eq. (S.36) involves a change of the sector due to the jump term. We can then write $\hat{A}_k = c_k^\dagger c_k$ in the new basis using Eq. (S.29) and then using Wick's theorem. Some algebra then yields

$$\langle \hat{c}_1^\dagger \mathbb{P}_k \hat{c}_k^\dagger \hat{c}_k \hat{c}_1 \rangle = \frac{4}{L^2} \left( n \sum_\lambda \frac{2\pi A_\lambda}{|e^{ik} - e^{i\lambda}|^2} - \frac{1}{L} \left| \sum_\lambda \frac{2\pi A_\lambda}{e^{ik} - e^{i\lambda}} \right|^2 + \frac{1}{L} \left| \sum_\lambda \frac{2\pi B_\lambda}{e^{ik} - e^{i\lambda}} \right|^2 \right). \tag{S.38}$$

For large $L$, sums can be replaced by integrals. To avoid a divergence in $\sum_\lambda A_\lambda/|e^{ik}-e^{i\lambda}|^2$, we can write

$$\sum_\lambda \frac{A_\lambda}{|e^{ik}-e^{i\lambda}|^2} = \sum_\lambda \frac{A_\lambda - A_k}{|e^{ik}-e^{i\lambda}|^2} + \sum_\lambda \frac{1}{|e^{ik}-e^{i\lambda}|^2} A_k = \sum_\lambda \frac{A_\lambda - A_k}{|e^{ik}-e^{i\lambda}|^2} + \frac{L^2}{4} A_k, \tag{S.39}$$

using the completeness relation, Eq. (S.31).

Finally we take the thermodynamic limit via the substitution $\sum_\lambda \cdot = \frac{L}{2\pi} \int_{-\pi}^\pi dk \cdot$. Putting everything together, we find

$$\begin{aligned}
F_A(k) = A_k - 2\pi \left( (A_k)^2 - \left| \frac{1}{\pi} \fint d\lambda \frac{A_\lambda}{e^{ik}-e^{i\lambda}} \right|^2 \right) + \frac{2n}{\pi} \fint d\lambda \frac{A_k - A_\lambda}{|e^{ik}-e^{i\lambda}|^2} \\
+ 2\pi \left( (B_k)^2 - \left| \frac{1}{\pi} \fint d\lambda \frac{B_\lambda}{e^{ik}-e^{i\lambda}} \right|^2 \right),
\end{aligned} \tag{S.40}$$

where the crossed integral denotes the Cauchy principal value of the integral. A similar analysis gives the expression for $F_B$:

$$F_B(k) = B_k - 2\pi \left( 2A_k B_k - \left[ \fint \frac{d\lambda}{\pi} \frac{A_\lambda}{e^{ik}-e^{i\lambda}} \fint \frac{d\lambda'}{\pi} \frac{B_\lambda}{e^{-ik}-e^{-i\lambda'}} + \text{c.c.} \right] \right) + \frac{2n}{\pi} \fint d\lambda \frac{B_k - B_\lambda}{|e^{ik}-e^{i\lambda}|^2}. \tag{S.41}$$

The above expressions can be written in a more suggestive form as follows. We first observe that

$$\begin{aligned}
\left| \fint \frac{d\lambda}{\pi} \frac{A_\lambda}{e^{ik}-e^{i\lambda}} \right|^2 &= \left( \frac{n}{2\pi} \right)^2 + \left| \fint \frac{d\lambda}{2\pi} A_\lambda \cot \frac{k-\lambda}{2} \right|^2, \\
\left| \fint \frac{d\lambda}{\pi} \frac{B_\lambda}{e^{ik}-e^{i\lambda}} \right|^2 &= \left| \fint \frac{d\lambda}{2\pi} B_\lambda \cot \frac{k-\lambda}{2} \right|^2,
\end{aligned} \tag{S.42}$$

where we have used the fact the fermion density is given by $n = \int dk A_k$, and $\int dk B_k = 0$ together with the identity $\frac{1}{1-e^{i\lambda}} = \frac{1}{2}\left(1 + i\cot\frac{\lambda}{2}\right)$. We can then write Eq. (S.40) as

$$F_A(k) = A_k + \frac{n^2}{2\pi} - 2\pi \left( (A_k)^2 - \left| \fint \frac{d\lambda}{2\pi} A_\lambda \cot\left( \frac{p-\lambda}{2} \right) \right|^2 - (A \to B) \right) + 2n \fint \frac{d\lambda}{2\pi} \frac{A_k - A_\lambda}{1-\cos(k-\lambda)}. \tag{S.43}$$

Furthermore, we note that

$$-\partial_p \cot\left( \frac{p-\lambda}{2} \right) = \frac{1}{1-\cos(p-\lambda)}. \tag{S.44}$$

The last term in Eq. (S.43) is the *Hadamard* finite part and can be written as

$$\fint \frac{d\lambda}{2\pi} \frac{A_p - A_\lambda}{1-\cos(p-\lambda)} = \partial_p \fint \frac{d\lambda}{2\pi} A_\lambda \cot\left( \frac{p-\lambda}{2} \right). \tag{S.45}$$

Putting these equations together and using the definition of the Hilbert transform, we recover the first line of Eq. (8) in the main text. One can similarly simplify the expression for $F_B(k)$. We note that

$$F_B(k) = B_k - 4\pi \left( A_k B_k - \fint \frac{d\lambda}{2\pi} A_\lambda \cot \frac{k-\lambda}{2} \fint \frac{d\lambda'}{2\pi} B_{\lambda'} \cot \frac{k-\lambda'}{2} \right) + 2n \fint \frac{d\lambda}{2\pi} \frac{B_k - B_\lambda}{1-\cos(k-\lambda)}, \tag{S.46}$$

and the last term can be written in a similar fashion to Eq. (S.45). One can then similarly recover the second line of Eq. (8) of the main text.

### 4. Analytic continuation

Next we consider the analytic continuation to the unit disk via

$$\mathcal{G}_A(z) = \int_0^{2\pi} \frac{dk}{2\pi} A_k \frac{e^{ik}+z}{e^{ik}-z} \quad \Rightarrow \quad \mathcal{G}_A(z) \text{ is analytic in } |z| < 1, \text{ and } \text{Re}\,\mathcal{G}_A(e^{ik-0^+}) = A_k. \tag{S.47}$$

A Kramers-Kronig relation follows as well:

$$\text{Im}\,\mathcal{G}_A(e^{ik-0^+}) = \fint_0^{2\pi} \frac{d\lambda}{2\pi} A(\lambda) \cot\left(\frac{k-\lambda}{2}\right). \tag{S.48}$$

We first consider Eq. (S.43) and focus on the expression in parentheses (considering just the $A$ term). Using the analytic properties of $\mathcal{G}_A(z)$, one can easily identify it as the real part of $(\mathcal{G}_A(z = e^{ip-0^+}))^2$. The $B$ term in parentheses can be treated identically. Finally, the last term, written in terms of the partial derivative, can be identified as the real part of $z\partial_z\mathcal{G}_A(z)$. One can similarly analytically continue the expression in Eq. (S.46) for $F_B(k)$. A similar argument shows that the expression in parentheses can be identified as the real part of the analytic function $\mathcal{G}_A(z)\mathcal{G}_B(z)$, and the last term in $F_B(k)$ can be dealt with similarly to the last term in $F_A(k)$. Putting all the different pieces together, we can then write

$$\frac{1}{\Gamma}\frac{d}{dt}\mathcal{G}_A(t,z) = -[\mathcal{G}_A + \frac{n^2}{2\pi} - 2\pi\mathcal{G}_A^2 + 2\pi\mathcal{G}_B^2 + 2nz\partial_z\mathcal{G}_A] + \mathcal{G}_{\Lambda_1}(z), \tag{S.49a}$$

$$\frac{1}{\Gamma}\frac{d}{dt}\mathcal{G}_B(t,z) = -[\mathcal{G}_B - 4\pi\mathcal{G}_A\mathcal{G}_B + 2nz\partial_z\mathcal{G}_B] + \mathcal{G}_{\Lambda_2}(z). \tag{S.49b}$$

Note that the Lagrange multiplier terms are also promoted to analytic functions corresponding to the functions $\Lambda_1(k) = \Lambda_k \sin k$ and $\Lambda_2(k) = -\Lambda_k(h - \cos k)$. The above equations can be simplified by defining

$$\mathcal{G}_{\pm}(z) = \mathcal{G}_A(z) \mp i\mathcal{G}_B(z). \tag{S.50}$$

Now combining the two equations in Eq. (S.49), we find

$$\frac{1}{\Gamma}\frac{d}{dt}\mathcal{G}_{\pm}(t,z) = -\left[\mathcal{G}_{\pm} + \frac{n^2}{2\pi} - 2\pi\mathcal{G}_{\pm}^2 + 2nz\partial_z\mathcal{G}_{\pm}\right] \pm ih\mathcal{G}_{\Lambda}(z) \mp i\mathcal{G}_{\Lambda_{\pm}}(z), \tag{S.51}$$

where we have defined $\Lambda_{\pm}(k) \equiv e^{\pm ik}\Lambda_k$. Note that, apart from the constraint, the equations for $\mathcal{G}_{\pm}$ are now decoupled from each other. Next we observe

$$\mathcal{G}_{\Lambda_{\pm}} = i\lambda_1 + z^{\pm 1}\mathcal{G}_{\Lambda}(z) \qquad \text{with} \quad \lambda_1 = \int dk \sin k \Lambda_k \tag{S.52}$$

where we have also used the fact that $\Lambda_k$ is odd in $k$ (while $A_k$ is even, and $B_k$ is also odd).

Next, we expand Eq. (S.51) to the zeroth order in powers of $z$. It turns out that both equations give the relation

$$\dot{n} = -n + \lambda_1, \qquad \text{where} \quad \lambda_1 = \int dk \Lambda(k) \sin k. \tag{S.53}$$

With this equation together with Eq. (S.52), we can write Eq. (S.51) as

$$\frac{1}{\Gamma}\frac{d}{dt}\left(\mathcal{G}_{\pm} \mp \frac{n}{2\pi}\right) = \pm\frac{n}{2\pi} - \left[\mathcal{G}_{\pm} + \frac{n^2}{2\pi} - 2\pi\mathcal{G}_{\pm}^2 + 2nz\partial_z\mathcal{G}_{\pm}\right] \pm i(h - z^{\pm 1})\mathcal{G}_{\Lambda}(z). \tag{S.54}$$

This motivates a change of variables for $\mathcal{G}_{\pm} \to \mathcal{G}_{\pm} \pm n/(2\pi)$. The above equation can be then written as

$$\frac{1}{\Gamma}\frac{d}{dt}\mathcal{G}_{\pm} = -\left[\mathcal{G}_{\pm} \mp 2n\mathcal{G}_{\pm} - 2\pi\mathcal{G}_{\pm}^2 + 2nz\partial_z\mathcal{G}_{\pm}\right] \pm i(h - z^{\pm 1})\mathcal{G}_{\Lambda}(z). \tag{S.55}$$

Thus, we recover the main result of the paper stated in Eq. (10) of the main text.

### 5. Multipole expansion

For a numerical solution, we consider a series expansion

$$A_k = \sum_{n=0}^{\infty} A_n \cos nk, \qquad B_k = \sum_{n=1}^{\infty} B_n \sin nk,$$

$$A_n = c_n \int_0^{2\pi} \frac{dk}{\pi} \cos(kn) A_k, \qquad B_n = \int_0^{2\pi} \frac{dk}{\pi} \sin(kn) B_k, \tag{S.56}$$

with the coefficient $c_n = 1/2$ for $n = 0$ and $c_n = 1$ for $n \geq 1$. The above series expansion corresponds to the multipole expansion of the analytic functions:

$$\mathcal{G}_A(z) = \sum_{n=0}^{\infty} A_n z^n, \qquad \mathcal{G}_B(z) = -i \sum_{n=1}^{\infty} B_n z^n. \tag{S.57}$$

Finally in terms of the analytic functions $\mathcal{G}_{\pm}$, we have

$$\mathcal{G}_+ = \sum_{n \geq 1}^{\infty} (A_n - B_n) z^n, \qquad \mathcal{G}_- = 2A_0 + \sum_{n \geq 1}^{\infty} (A_n + B_n) z^n, \quad \text{with} \quad A_0 = \frac{n}{2\pi}. \tag{S.58}$$

Finally we point out that the coefficients $A_n, B_n$ are constrained by Eq. (S.34)

$$\begin{aligned}
A_0 - \frac{1}{4\pi} &= -\sum_{m=1}^{\infty} B_{2m} + h \sum_{m=0}^{\infty} B_{2m+1}, \\
A_n &= -B_n + \sum_{\substack{m > n \\ m+n \text{ even}}}^{\infty} -2B_m + \sum_{\substack{m > n \\ m+n \text{ odd}}}^{\infty} 2h B_m, \qquad n \geq 1.
\end{aligned} \tag{S.59}$$

To numerically solve these equations, we first eliminate the function $\mathcal{G}_\Lambda(z)$ in Eq. (9) to obtain (setting $\Gamma = 1$):

$$\frac{\mathrm{d}}{\mathrm{d}t}[(-h + \frac{1}{z})\mathcal{G}_+ - (h - z)\mathcal{G}_-] = -(-h + \frac{1}{z})[1 - 2\pi\mathcal{G}_+ - 2n + 2nz\partial_z]\mathcal{G}_+ + (h - z)[1 - 2\pi\mathcal{G}_- + 2n + 2nz\partial_z]\mathcal{G}_-. \tag{S.60}$$

Next, with the aid of the condition in Eq. (S.59), Eq. (S.60) is then expanded in powers of $z$ to find the coefficients $B_n$ which can be used to write all non-vanishing fermionic correlation functions. More precisely, we can write the correlations between two fermions at a distance $l$ using the inverse Fourier transformation

$$\left\langle \hat{c}_i^\dagger \hat{c}_{i+l} \right\rangle = \frac{1}{2\pi} \int_0^{2\pi} dk e^{ikl} \left\langle \hat{c}_k^\dagger \hat{c}_k \right\rangle = \int_0^{2\pi} dk \cos kl \ [A_0 + \sum_{n=1}^{\infty} A_m \cos kn] = 2\pi A_0 \delta_{l,0} + \pi \sum_{n=1}^{\infty} A_n \delta_{|l|,n}. \tag{S.61}$$

Similarly,

$$\langle \hat{c}_i \hat{c}_{i+l} \rangle = \text{sgn}(l) \ \pi \sum_{n=1}^{\infty} B_n \delta_{|l|,n}. \tag{S.62}$$

We solve Eq. (S.60) by time-evolving $B_n$ starting from an infinite temperature state $[A_k(t=0) = 1/4\pi, \ B_k(t=0) = 0]$ and keeping up to $n_{\max} = 80$ terms in the expansion.

## S.V. Spin correlations from fermions

Under the Jordan-Wigner transformation defined by $\sigma_j^- = e^{i\pi \sum_{i \geq 1}^j \hat{c}_i^\dagger \hat{c}_i} \hat{c}_j$, we can write the spin correlator $\langle \sigma_i^x \sigma_j^x \rangle$ (assuming $j > i$) as

$$\begin{aligned}
\langle \sigma_i^x \sigma_j^x \rangle &= \left\langle \hat{N}_i (\prod_{i < l < j} \hat{M}_l \hat{N}_l) \hat{M}_j \right\rangle \\
&= \left\langle \hat{N}_i \hat{M}_{i+1} \hat{N}_{i+1} ... \hat{M}_{j-1} \hat{N}_{j-1} \hat{M}_j \right\rangle,
\end{aligned} \tag{S.63}$$

where we define

$$\hat{M}_i \equiv \hat{c}_i^\dagger + \hat{c}_i, \qquad \hat{N}_i \equiv \hat{c}_i^\dagger - \hat{c}_i. \tag{S.64}$$

as well as the fermionic correlations

$$
\begin{aligned}
S_{i,j} &\equiv \left\langle \hat{N}_i \hat{N}_j \right\rangle, \\
Q_{i,j} &\equiv \left\langle \hat{M}_i \hat{M}_j \right\rangle, \\
G_{i,j} &\equiv \left\langle \hat{N}_i \hat{M}_j \right\rangle.
\end{aligned}
\tag{S.65}
$$

Since both function $A_k$ and $B_k$ are real, one can see that the correlation functions $Q_{i,j} = -S_{i,j} = \delta_{i,j}$ for any $i$ and $j$. The spin correlations at a distance $R$ are then given by the determinant

$$
\left\langle \sigma_i^x \sigma_{i+R}^x \right\rangle =
\begin{vmatrix}
G_{-1} & G_{-2} & . & . & . & . & .... & G_{-R} \\
G_0 & G_{-1} & . & . & . & . & .... & G_{-R+1} \\
 & 0 & . & . & . & . & & \\
 & . & . & . & & & & . \\
 & . & . & . & & & & . \\
G_{R-2} & G_{R-3} & . & . & . & . & ... & G_{-1}
\end{vmatrix},
\tag{S.66}
$$

where

$$
G_l = 2\,\mathrm{Re}\left\{ \left\langle \hat{c}_i^\dagger \hat{c}_{i+l} \right\rangle - \left\langle \hat{c}_i \hat{c}_{i+l} \right\rangle \right\} - \delta_{l,0}.
\tag{S.67}
$$

Alternatively, using the expressions Eqs. (S.61) and (S.62), we have

$$
\begin{aligned}
G_0 &= 4\pi A_0 - 1, \\
G_l &= 2\pi (A_{|l|} - \mathrm{sgn}(l) B_{|l|}) \qquad \text{for} \quad l \neq 0.
\end{aligned}
\tag{S.68}
$$

### S.VI. Contrast against free fermions & quench dynamics

In this section, we provide a solution to the quadratic fermionic problem where the Lindblad operators take the form of independent fermionic losses, $\hat{L}_i = \hat{c}_i$, can be solved exactly assuming periodic boundary conditions $\hat{c}_{L+1} = \hat{c}_1$; our presentation closely follows Ref. [S10]. Applying a Fourier transformation to $H_o$ in Eq. (S.22), we get

$$
\hat{H}_o = 2\sum_{k>0}(h - \cos k)[\hat{c}_k^\dagger \hat{c}_k + \hat{c}_{-k}^\dagger \hat{c}_{-k}] - i\sin k[\hat{c}_k \hat{c}_{-k} + \hat{c}_k^\dagger \hat{c}_{-k}^\dagger].
\tag{S.69}
$$

For a single momentum mode, the basis is spanned by the states

$$
\begin{aligned}
|\phi_1^k\rangle &= |0,0\rangle, \\
|\phi_2^k\rangle &= \hat{c}_k^\dagger |0,0\rangle = |k,0\rangle, \\
|\phi_3^k\rangle &= \hat{c}_{-k}^\dagger |0,0\rangle = |0,-k\rangle, \\
|\phi_4^k\rangle &= \hat{c}_{-k}^\dagger \hat{c}_k^\dagger |0,0\rangle = |k,-k\rangle,
\end{aligned}
\tag{S.70}
$$

where we can write the Hamiltonian on the form $\hat{H}_o = \sum_{k>0} \hat{h}_k$, as we define the operators

$$
\hat{c}_k = \begin{pmatrix} 0 & 1 & 0 & 0 \\ 0 & 0 & 0 & 0 \\ 0 & 0 & 0 & -1 \\ 0 & 0 & 0 & 0 \end{pmatrix}, \quad
\hat{c}_{-k} = \begin{pmatrix} 0 & 0 & 1 & 0 \\ 0 & 0 & 0 & 1 \\ 0 & 0 & 0 & 0 \\ 0 & 0 & 0 & 0 \end{pmatrix}, \quad
\hat{h}_k = 2\begin{pmatrix} -(h-\cos k) & 0 & 0 & -i\sin k \\ 0 & 0 & 0 & 0 \\ 0 & 0 & 0 & 0 \\ i\sin k & 0 & 0 & (h-\cos k) \end{pmatrix}.
\tag{S.71}
$$

Since the modes $k$ are decoupled, we can then write a Lindblad master equation for each mode

$$
\begin{aligned}
\frac{d\hat{\rho}_k}{dt} &= -i\left[\hat{h}_k, \hat{\rho}_k\right] + \mathcal{D}_k(\hat{\rho}_k), \\
\mathcal{D}_k(\hat{\rho}_k) &= \Gamma(\hat{c}_k \hat{\rho}_k \hat{c}_k^\dagger - \frac{1}{2}\left\{\hat{c}_k^\dagger \hat{c}_k, \hat{\rho}_k\right\} + \hat{c}_{-k} \hat{\rho}_k \hat{c}_{-k}^\dagger - \frac{1}{2}\left\{\hat{c}_{-k}^\dagger \hat{c}_{-k}, \hat{\rho}_k\right\}).
\end{aligned}
\tag{S.72}
$$

In the steady state, we find the only non-vanishing elements of the density matrix to be

$$\hat{\rho}_k(t \to \infty) = \begin{pmatrix} \rho_{11} & 0 & 0 & \rho_{14} \\ 0 & \rho_{22} & 0 & 0 \\ 0 & 0 & \rho_{33} & 0 \\ \rho_{14}^* & 0 & 0 & \rho_{44} \end{pmatrix}, \tag{S.73}$$

where, with the aid of the condition $\text{Tr}\{\rho_k\} = 1$, we find

$$A_k = \frac{1}{2\pi}(\rho_{22} + \rho_{44}) = \frac{1}{4\pi} \frac{\sin^2 k}{1 + h^2 - 2h\cos k + (\frac{\Gamma}{4})^2}, \tag{S.74a}$$

$$B_k = \frac{1}{2\pi i}\rho_{14}^* = \frac{-1}{4\pi} \frac{\sin k(h - \cos k + \frac{i\Gamma}{4})}{1 + h^2 - 2h\cos k + (\frac{\Gamma}{4})^2}. \tag{S.74b}$$

Alternatively, in the limit $\Gamma \to 0$, we can simply drop the nonlinear terms in Eq. (S.35) and just keep the leading linear term in $F_{A,B}$. Then the steady state is described by

$$\frac{dA_k}{dt} = -A_k + \Lambda_k \sin\theta_k = 0, \tag{S.75a}$$

$$\frac{dB_k}{dt} = -B_k - \Lambda_k \cos\theta_k = 0, \tag{S.75b}$$

which then yields

$$A_k \cos\theta_k + B_k \sin\theta_k = 0. \tag{S.76}$$

Together with the constraint in Eq. (S.33), we can solve the two equations for $A_k$, and $B_k$ in the steady state which yields

$$A_k = \frac{1}{2}\sin^2\theta_k, \qquad B_k = -\frac{1}{2}\sin\theta_k\cos\theta_k. \tag{S.77}$$

Upon substituting for $\theta_k$ in terms of $h$, we find

$$A_k = \frac{1}{4\pi} \frac{\sin^2 k}{1 + h^2 - 2h\cos k}, \tag{S.78a}$$

$$B_k = \frac{-1}{4\pi} \frac{\sin k(h - \cos k)}{1 + h^2 - 2h\cos k}. \tag{S.78b}$$

These equations simply reproduce Eq. (S.78) in the limit $\Gamma \to 0$.

We can also find these expressions in terms of the occupation of the Bogoliubov fermions. Recalling Eq. (S.32), we can identify the occupation number of Bogoliubov fermions as

$$N_k = \frac{1}{2} + p_k = \frac{1}{2}(1 - \cos\theta_k) \tag{S.79}$$

It turns out that the result is identical to the ground state corresponding to $h_0 = \infty$ (corresponding to a state will all spins pointing up) to $h$ [S11, S12]. The correlation function in this model is described by [S11]

$$C_r = \begin{cases} 2^{-r}\cos[r\arccos h], & h \le 1 \\ (2h)^{-r} & h \ge 1 \end{cases} \tag{S.80}$$

Therefore, the correlation length is given by

$$\xi = \begin{cases} 1/\log 2, & h \le 1 \\ 1/\log(2h) & h \ge 1 \end{cases} \tag{S.81}$$

While the correlation length does not change below the critical point, we note the oscillatory nature of the correlation function. In some sense, the correlations are *best formed* at the critical point [S11].

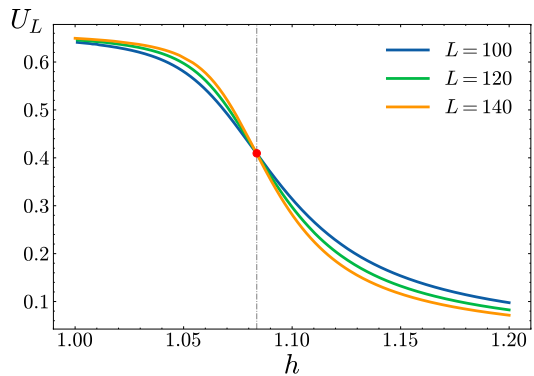

FIG. S.2. The Binder cumulant $U_L$ calculated in the ground state of $\hat{H}_2$ with $J_2 = 0.05$ for multiple system sizes. The red dot marks, $h_c$, the point where the curves cross.

## S.VII. Quantum Critical Point for Non-Integrable Hamiltonians

For the non-integrable next-nearest-neighbor Ising model considered in Fig. 3 in the main text, we resort to finding the quantum critical point, $h_c$, by calculating the Binder cumulant [S13] in the ground state

$$U_L = 1 - \frac{\langle S_x^4 \rangle_L}{3 \langle S_x^2 \rangle_L^2}, \tag{S.82}$$

where $S_x = \sum_{i=1}^{L} \sigma_i^x$ is the total magnetization operator on the chain. The ground state is obtained using standard MPS techniques for $100 \leq L \leq 140$. The critical point $h_c$ is where the different curves for large system sizes.

---



\end{document}